\documentclass[conference]{IEEEtran}
\ifCLASSINFOpdf
\else
\fi
\usepackage{setspace}
\usepackage{verbatim}
\usepackage{cite}
\usepackage{amsmath,amssymb,amsfonts,amsthm}
\usepackage[top=0.7in, bottom=1.03in, left=0.71in, right=0.71in]{geometry}
\usepackage{algorithm}
\usepackage{graphicx}
\usepackage{comment}
\usepackage{textcomp}
\usepackage{tikz}
\usepackage{color}
\usepackage{stfloats}

\def\BibTeX{{\rm B\kern-.05em{\sc i\kern-.025em b}\kern-.08em
    T\kern-.1667em\lower.7ex\hbox{E}\kern-.125emX}}
\usepackage{bm}      
\usepackage{multirow}
\usepackage{algpseudocode}
\usepackage{amsmath,amssymb}
\usepackage{tikz}
\usetikzlibrary{positioning,calc}
\begin{document}

\title{
Joint Movement and Compression Ratio Design for Mobile Embodied AI Networks (MEAN)
}

\author{
\IEEEauthorblockN{
Yahao Ding\IEEEauthorrefmark{1},
Jiaxiang Wang\IEEEauthorrefmark{1},
Zhouxiang Zhao\IEEEauthorrefmark{2},
Zhaohui Yang\IEEEauthorrefmark{2},
Mingzhe Chen\IEEEauthorrefmark{3},
Mohammad Shikh-Bahaei\IEEEauthorrefmark{1}
}

\IEEEauthorblockA{
\IEEEauthorrefmark{1}King's College London,
\IEEEauthorrefmark{2}Zhejiang University,
\IEEEauthorrefmark{3}University of Miami\\
Email:
\IEEEauthorrefmark{1}\{yahao.ding, jiaxiang.wang, m.sbahaei\}@kcl.ac.uk,
\IEEEauthorrefmark{2}\{zhouxiangzhao, yang\_zhaohui\}@zju.edu.cn,\\
\IEEEauthorrefmark{3}mingzhe.chen@miami.edu
}
}

\maketitle
\begin{abstract}
Mobile embodied AI networks (MEAN) enable embodied agents to perceive, reason, communicate, and act in wireless environments. In such networks, agent mobility can improve channel conditions, while semantic compression can reduce transmission payloads. However, movement consumes energy, and stronger compression incurs additional computational cost. This paper studies joint movement, semantic compression, and transmit power design for an uplink MEAN system. We formulate a max-min energy efficiency (EE) problem by jointly optimizing transmit power, movement distance, and semantic compression ratio under controllable power constraints. The problem is non-convex due to the coupled signal-to-interference-plus-noise ratio (SINR), mobility-dependent channel gains, and fractional EE objective. To solve it, we propose an alternating optimization (AO)-Dinkelbach algorithm, where the fractional objective is handled by the Dinkelbach transformation, transmit power is updated via successive convex approximation (SCA), and movement distance is updated by coordinate-wise grid search. Simulation results show that the proposed scheme outperforms no-mobility and no-compression baselines, demonstrating the benefit of jointly exploiting mobility control, semantic compression, and power allocation in MEAN.
\end{abstract}

\section{Introduction}
Mobile embodied AI networks (MEAN) are emerging as a key use case for future wireless networks~\cite{zhang2025embodiedlargeai}, where agents such as mobile robots, autonomous vehicles, drones, and edge AI devices need to perceive physical environments, reason over task-relevant information, communicate with network infrastructure, and act through physical movement. Unlike conventional wireless terminals, embodied AI agents can adapt both their communication strategies and physical locations according to task requirements and channel conditions. This motivates the concept of MEAN, in which communication, computation, and mobility are jointly coordinated to support embodied decision-making and task execution.


Semantic communication is well suited to MEAN, as embodied agents can transmit task-relevant semantic features or compressed tokens instead of raw sensory data, thereby reducing the effective transmission payload. Existing studies have shown the benefits of deep-learning-enabled, context-aware, and task-oriented semantic communication in bandwidth-limited and noisy wireless environments~\cite{11675865,xie2021deep,Gunduz2023Beyond,xie2022task,11432269}. However, most existing designs assume fixed transmitter locations and do not exploit agent mobility as a controllable communication resource. Mobility control can reduce path loss and improve uplink rates, as shown in mobility-aware and UAV communication studies~\cite{zeng2016wireless,zeng2019accessing,11216397}, but it also incurs locomotion energy consumption. Meanwhile, stronger semantic compression reduces transmission load but increases computational cost. Therefore, semantic compression and mobility control are inherently coupled in MEAN, motivating a unified design that jointly considers semantic compression, mobility control, and communication resource allocation.

\subsection{Related Works}
Recent studies have investigated semantic communication and deep joint source-channel coding (Deep-JSCC) for wireless image, text, and task-oriented transmission~\cite{bourtsoulatze2019deep,xie2021deep,yang2023swinjscc,xie2022task,ma2023task}. These works develop end-to-end neural encoders or task-oriented semantic transmission schemes to improve robustness and reduce communication overhead. However, they mainly focus on semantic encoding, source-channel coding, or task performance under static transmitter locations, and thus do not exploit the mobility of embodied agents as an additional design dimension for improving wireless channel conditions.

In parallel, joint communication and computation optimization for mobile edge computing (MEC) has been extensively studied in~\cite{mao2016dynamic,mach2017mobile,you2017energy}. These works provided foundational frameworks for computation offloading and energy-efficient resource allocation under communication and computation constraints. However, most MEC-oriented works focus on the communication-computation trade-off and treat agent locations as fixed. They therefore do not capture the mobility-aware communication-computation coupling that arises in MEAN, where physical movement directly affects both wireless channel quality and energy consumption.

Meanwhile, mobility-aware communication and UAV-assisted networks have been studied in~\cite{zeng2016wireless,zeng2019accessing}. These works showed that trajectory and placement optimization can improve coverage, throughput, and energy efficiency by adapting wireless link geometry. Recent embodied AI studies also envision autonomous agents that perceive, reason, communicate, and act in resource-constrained edge environments~\cite{zhang2025toward}. 
Nevertheless, existing studies mainly consider semantic compression, computation offloading, or mobility control separately. To the best of our knowledge, joint optimization of semantic compression ratio, movement distance, and transmit power for max-min energy efficiency in MEAN has not been fully investigated, which is the focus of this paper.



\subsection{Contributions}

Motivated by these challenges, this paper investigates joint movement, semantic compression, and power allocation for uplink MEAN, where each embodied agent jointly optimizes its transmit power, movement distance, and semantic compression ratio to improve the max-min energy efficiency (EE). The main contributions are summarized as follows:

\begin{itemize}
    \item  
    We propose a unified system model for MEAN, where each mobile embodied agent jointly determines its semantic compression ratio, transmit power, and movement distance. The integrated communication-computation-mobility framework captures the compression-aware effective rate that scales inversely with the compression ratio, the mobility-induced channel gain variation through reduced path loss, and the aggregate energy consumption from transmission, semantic computation, physical locomotion, and fixed circuit operation.
    
    \item
    We formulate a max-min EE optimization problem that maximizes the minimum per-agent EE, thereby enforcing fairness among heterogeneous mobile embodied agents. The formulated problem jointly optimizes semantic compression ratio, transmit power, and movement distance under controllable power constraints. It is a non-convex fractional program with coupled optimization variables through the minimum mean square error (MMSE) combining and signal-to-interference-plus-noise ratio (SINR) expressions.
    
    \item 
    To tackle this non-convex fractional problem, we propose a two-layer iterative algorithm. The outer layer uses the Dinkelbach transformation to convert the fractional objective into a parametric subtractive form. The inner layer adopts alternating optimization (AO) with successive convex approximation (SCA), where semantic compression is updated via one-dimensional search, transmit power is optimized by linearizing the interference term through first-order Taylor approximation, and movement distance is updated by coordinate-wise grid search.

    \item 
    Simulation results demonstrate that the proposed optimization framework significantly improves the max-min EE compared with baseline schemes that disable either mobility control or semantic compression, validating the importance of jointly exploiting agent mobility and semantic compression in MEAN.
\end{itemize}

The rest of this paper is organized as follows. Section~II presents the system model. Section~III formulates the max-min EE optimization problem. Section~IV develops the proposed algorithm. Section~V provides simulation results. Section~VI concludes this paper.

\section{System Model}
\label{sec:system}

As shown in Fig.~1, we consider a single-cell uplink MEAN system modeled as a multi-user single-input multiple-output (MU-SIMO) system, where \(K\) single-antenna mobile agents simultaneously transmit to a base station (BS) with \(M\) receive antennas. Let $\mathcal{K} = \{1, \ldots, K\}$ denote the set of agents. Each agent $k$ is initially located at distance $D_k$ from the BS and can move toward the BS by a controllable displacement $d_k \in [d_k^{\min}, d_k^{\max}]$, yielding the post-movement communication distance $r_k(d_k) = D_k - d_k$ with $D_k > d_k^{\max}$.

\begin{figure}
    \centering
    \includegraphics[width=\linewidth]{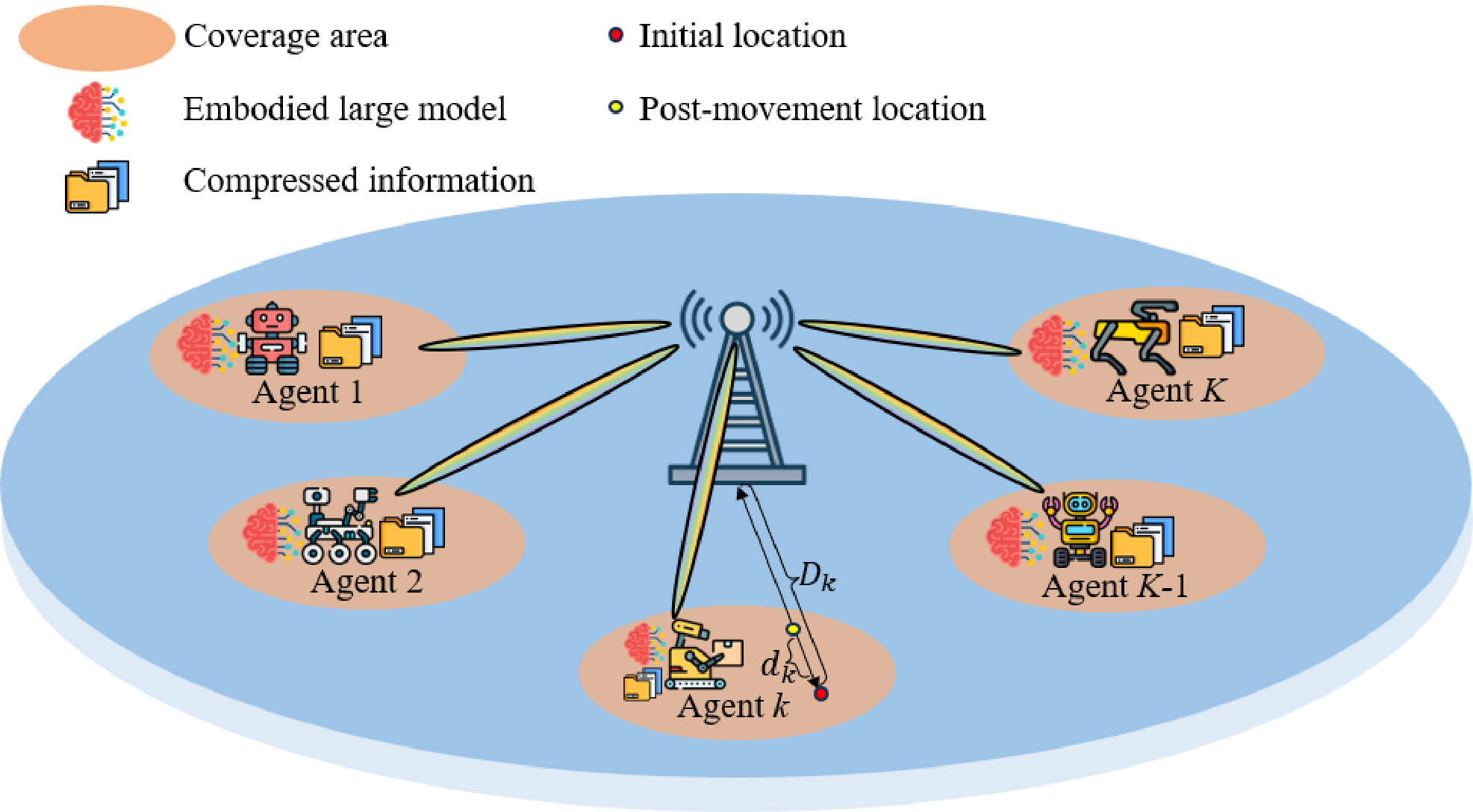}
    \caption{Illustration of the considered uplink MEAN system.}
    \label{fig: system model}
    \vspace{-1.5em}
\end{figure}
 \vspace{-0.8em}
\subsection{Channel Model}
We adopt a large-scale fading channel model, and the uplink channel vector from agent $k$ to the BS is modeled as
\begin{equation}
\bm{h}_k(d_k) = \sqrt{\beta_k(d_k)} \, \bm{g}_k, \quad \bm{g}_k \sim \mathcal{CN}(\bm{0}, \bm{I}_M),
\end{equation}
where $\beta_k(d_k)$ captures the distance-dependent large-scale channel gain and $\bm{g}_k \in \mathbb{C}^{M \times 1}$ represents the small-scale fading vector. A commonly used path-loss model is given by
$\beta_k(d_k) = \beta_0 (D_k - d_k)^{-\delta}$,
where $\beta_0$ is the reference channel gain at unit distance and $\delta > 2$ is the path-loss exponent.
\vspace{-0.5em}
\subsection{Uplink Transmission and MMSE Receive Combining}
During the uplink transmission phase, each agent $k$ transmits a symbol $s_k$ with unit average power, i.e., $\mathbb{E}[|s_k|^2] = 1$. The transmitted signal from agent $k$ is $x_k = \sqrt{p_k} s_k$, where $p_k \geq 0$ denotes the transmit power. Given the system bandwidth $B$, the received signal at the BS is expressed as:
\vspace{-0.5em}
\begin{equation}
\bm{y} = \sum_{k=1}^{K} \sqrt{p_k} \, \bm{h}_k(d_k) s_k + \bm{n},
\end{equation}
where $\bm{n} \sim \mathcal{CN}(\bm{0}, \sigma^2 \bm{I}_M)$ is the additive white Gaussian noise (AWGN) with $\sigma^2 = N_0 B$ denoting the noise power and $N_0$ being the noise power spectral density.
The BS employs MMSE receive combining to detect each agent's symbol. Specifically, the combining vector for detecting the symbol of agent $k$ is given by:
\vspace{-0.5em}
\begin{equation}
\bm{w}_k = \left( \sum_{j=1}^{K} p_j \bm{h}_j(d_j) \bm{h}_j^H(d_j) + \sigma^2 \bm{I}_M \right)^{-1} \sqrt{p_k} \, \bm{h}_k(d_k).
\label{eq:mmse}
\end{equation}
The detected symbol for agent $k$ is obtained as $\hat{s}_k = \bm{w}_k^H \bm{y}$. The post-combining SINR of agent $k$ is:
\vspace{-0.5em}
\begin{equation}
\gamma_k = \frac{p_k \left| \bm{w}_k^H \bm{h}_k(d_k) \right|^2}{\sum_{j \neq k} p_j \left| \bm{w}_k^H \bm{h}_j(d_j) \right|^2 + \sigma^2 \| \bm{w}_k \|^2}.
\label{SINR}
\end{equation}
The achievable uplink rate of agent $k$ is thus given by
\begin{equation}
R_k = B \log_2 (1 + \gamma_k).
\label{Rate}
\end{equation}

\subsection{Semantic Compression Model}

On the source side, each mobile embodied agent uses an embodied AI model to extract task-relevant semantic information from its sensory observations. Instead of transmitting raw sensory data, agent \(k\in\mathcal{K}\) transmits a compressed semantic representation with compression ratio \(\rho_k \in [\rho_k^{\min},\rho_k^{\max}]\), where a smaller \(\rho_k\) indicates stronger compression and fewer transmitted semantic tokens.

Following a raw-information-equivalent effective rate model, the compression-aware effective rate of agent \(k\) is defined as
\begin{equation}
R_k^{\mathrm{eff}} = \frac{R_k}{\rho_k},
\end{equation}
where \(R_k^{\mathrm{eff}}\) denotes the equivalent rate for delivering the original uncompressed semantic information. Thus, semantic compression can improve the effective information delivery rate under a given physical uplink rate, but it also incurs additional computational cost.

\subsection{Energy Consumption Model}
Based on the above system model, the total power consumption of agent $k$ includes the following four components: 1) the uplink transmit power $p_k$; 2) the token compression power consumption, modeled as $A_k \ln(\frac{1}{\rho_k})$, where $A_k > 0$ is the compression coefficient determined by the computational complexity of the compression algorithm; 3) the mobility power consumption, modeled as $\frac{F_k d_k}{T_k}$, where $F_k$ is the mobility energy coefficient and $T_k$ denotes the motion duration; and 4) the fixed circuit and baseline processing power $P_k^{\mathrm{c}}$. Therefore, the total power consumption of agent $k$ is:
\begin{equation}
P_k^{\mathrm{tot}} = p_k + A_k \ln\left( \frac{1}{\rho_k} \right) + \frac{F_k d_k}{T_k}+P_k^{\mathrm{c}}.
\label{P_total}
\end{equation}

Accordingly, the EE of agent $k$ is defined as
\begin{equation}
\eta_k 
= 
\frac{R_k^{\mathrm{eff}}}{P_k^{\mathrm{tot}}}
=
\frac{
\frac{B}{\rho_k}\log_2\!\bigl(1+\gamma_k\bigr)
}{
p_k 
+ A_k \ln\!\left(\frac{1}{\rho_k}\right) 
+ \frac{F_k d_k}{T_k}
+ P_k^{\mathrm{c}}
}.
\label{eq:EE}
\end{equation}

Since $P_k^{\mathrm{c}}$ is fixed and cannot be controlled by the resource allocation algorithm, we further define the controllable power consumption of agent $k$ as
\begin{equation}
P_k^{\mathrm{ctrl}}
=
p_k 
+ A_k \ln\left( \frac{1}{\rho_k} \right) 
+ \frac{F_k d_k}{T_k}.
\label{eq:P_ctrl}
\end{equation}
In the following optimization problem, $P_k^{\mathrm{ctrl}}$ is constrained by the controllable power budget $P_k^{\max}$, while the total power $P_k^{\mathrm{tot}}=P_k^{\mathrm{ctrl}}+P_k^{\mathrm{c}}$ is used in the EE denominator.

\section{Problem Formulation}
\label{sec:problem}
In this section, we formulate the EE maximization problem for the considered MU-SIMO uplink system. To ensure fairness among heterogeneous agents, we adopt a max--min design criterion that maximizes the minimum per-agent EE.

Let $\bm{\rho} \triangleq [\rho_1,\ldots,\rho_K]^T$, $\bm{p} \triangleq [p_1,\ldots,p_K]^T$, and $\bm{d} \triangleq [d_1,\ldots,d_K]^T$ denote the vectors of semantic compression ratios, transmit powers, and movement distances, respectively. Since the optimal receive combining vectors $\{\bm{w}_k\}_{k\in\mathcal{K}}$ at the BS are uniquely determined by $\{\bm{p}, \bm{d}\}$ via the MMSE solution in \eqref{eq:mmse}, they are not treated as independent optimization variables. The joint optimization problem is formulated as
\begin{subequations}\label{prob:P1}
\begin{align}
(\mathcal{P}1): \; \max_{\bm{\rho}, \bm{p}, \bm{d}} \; & \min_{k \in \mathcal{K}} \; \eta_k \label{prob:P1_obj} \tag{\theequation} \\
\text{s.t.} \quad 
&  P_k^{\mathrm{ctrl}} \leq P^{\max}_k, \; \forall k \in \mathcal{K},\label{prob:P1a} \\
& p_k\geq 0, \; \forall k \in \mathcal{K}, \label{prob:P1b} \\
& d_k^{\min} \leq d_k \leq d_k^{\max}, \; \forall k \in \mathcal{K}, \label{prob:P1c}\\
& \rho_k^{\min} \leq \rho_k \leq \rho_k^{\max}, \; \forall k \in \mathcal{K}, \label{prob:P1d} 
\end{align}
\end{subequations}
where $P_k^{\max}$ denotes the maximum controllable power budget of agent $k$. 
Constraint \eqref{prob:P1a} restricts the controllable power consumption, including transmission, semantic compression, and mobility power. Constraints \eqref{prob:P1b}-\eqref{prob:P1d} respectively specify the transmit-power non-negativity constraint, movement-distance bounds, and compression-ratio bounds for all agents.



\vspace{-0.2em}

\section{Proposed Solution Algorithm}


Problem $(\mathcal{P}1)$ is a non-convex max-min fractional programming problem. 
The non-convexity comes from the fractional EE objective, the coupled SINR expression, and the mobility-dependent channel model. To tackle this, we propose a two-layer iterative 
algorithm: the outer layer applies Dinkelbach's method to handle the 
fractional objective, while the inner layer employs alternating optimization (AO) 
with successive convex approximation (SCA) to resolve the non-convexity.


\subsection{Dinkelbach Transformation}
To handle the fractional max-min objective,  we introduce an auxiliary parameter \(\lambda \geq 0\) and apply the Dinkelbach transformation. For a given \(\lambda^{(n)}\) at the \(n\)-th outer iteration, the original problem is transformed into the following form:
\vspace{-0.5em}
\begin{subequations}\label{prob:P2}
\begin{align}
(\mathcal{P}2): \; \max_{\bm{\rho}, \bm{p}, \bm{d},\, \tau} \; & \tau 
\tag{\theequation}\label{prob:P2_obj}\\
\text{s.t.} \quad 
& R_k^{\mathrm{eff}} - \lambda^{(n)} P_k^{\mathrm{tot}} \geq \tau,
\quad \forall k\in\mathcal{K}, \label{prob:P2a}\\
& \eqref{prob:P1a}\text{--}\eqref{prob:P1d}. \label{prob:P2b}
\end{align}
\end{subequations}
where $\tau$ represents the minimum subtractive utility among all agents. For a fixed $\lambda^{(n)}$, maximizing \(\tau\) improves the minimum value of \(R_k^{\mathrm{eff}}-\lambda^{(n)} P_k^{\mathrm{tot}}\). After solving \((\mathcal{P}2)\) at the \(n\)-th outer iteration, the Dinkelbach parameter is updated by
\begin{equation}
\lambda^{(n+1)}
=
\min_{k\in\mathcal{K}}
\frac{R_k^{\mathrm{eff}}(\bm{p}^{(n)},\bm{d}^{(n)},\boldsymbol{\rho}^{(n)})}
{P_k^{\mathrm{tot}}(\bm{p}^{(n)},\bm{d}^{(n)},\bm{\rho}^{(n)})}.
\label{lamada}
\end{equation}
The outer iteration terminates when
$\left|\lambda^{(n+1)}-\lambda^{(n)}\right| \leq \epsilon
$, where \(\epsilon\) is a predefined convergence threshold.


\subsection{Compression Ratio Update}

For fixed transmit power \(\boldsymbol{p}\) and movement distance \(\boldsymbol{d}\), the achievable rate \(R_k\) of each agent is fixed. Therefore, the compression ratio of each agent can be updated independently. The subproblem for \(\rho_k\) is given by
\begin{equation}
\max_{\rho_k}
\frac{R_k}{\rho_k}
-
\lambda
\left(
P_k^{\mathrm{ctrl}} + P_k^c
\right),
\end{equation}
subject to the compression ratio constraint in \eqref{prob:P1d} and the controllable power constraint in \eqref{prob:P1a}.

From \(P_k^{\mathrm{ctrl}} \leq P_k^{\max}\), the compression ratio must satisfy
\begin{equation}
A_k \ln\left(\frac{1}{\rho_k}\right)
\leq
P_k^{\max}
-
p_k
-
\frac{F_k d_k}{T_k}.
\end{equation}
Thus, the feasible lower bound of \(\rho_k\) is given by
\begin{equation}
\rho_k^{\mathrm{lb}}
=
\max
\left\{
\rho_k^{\min},
\exp\left(
-
\frac{
P_k^{\max}
-
p_k
-
\frac{F_k d_k}{T_k}
}{
A_k
}
\right)
\right\}.
\end{equation}
The feasible upper bound is $
\rho_k^{\mathrm{ub}}=\rho_k^{\max}$.

The compression-ratio subproblem is one-dimensional and can be efficiently solved by direct search over the compact feasible interval $[\rho_k^{\mathrm{lb}},\rho_k^{\mathrm{ub}}]$. Specifically, the update is given by

\begin{equation}
\rho_k^\star
=
\arg\max_{\rho_k \in [\rho_k^{\mathrm{lb}},\rho_k^{\mathrm{ub}}]}
\left(
R_k^{\mathrm{eff}}
-
\lambda P_k^{\mathrm{tot}}
\right).
\label{rho_update}
\end{equation}

\subsection{Transmit Power Update via SCA}
For fixed $\bm{d}$ and $\boldsymbol{\rho}$, the receive combining vectors are first updated according to the MMSE rule in \eqref{eq:mmse}. 
Although the achievable rate has been defined in \eqref{Rate}, the transmit power update remains non-convex because the SINR in \eqref{SINR} contains interference terms coupled across agents. 
To obtain a tractable approximation, we rewrite the achievable rate into a difference-of-concave form.

For notational simplicity, we define
\begin{equation}
a_{k,j}
=
\left|
\bm{w}_k^{H}\bm{h}_j(d_j)
\right|^2,
\qquad
\nu_k
=
\sigma^2\|\bm{w}_k\|^2 .
\end{equation}
Then, based on (4) and (5), the achievable rate of agent $k$ can be equivalently written as
\begin{equation}
R_k
=
B\log_2
\left(
\sum_{j\in\mathcal{K}} a_{k,j}p_j+\nu_k
\right)
-
B\log_2
\left(
\sum_{j\neq k} a_{k,j}p_j+\nu_k
\right).
\label{eq:rate_dc}
\end{equation}
This expression is non-convex in $\bm{p}$ because it is the difference of two concave logarithmic functions. Therefore, to construct a convex approximation, we apply SCA to the second logarithmic term by using the first-order Taylor expansion. At the current local point $\bm{p}^{(r)}$, we define the interference-plus-noise term as
\begin{equation}
\Omega_k^{(r)}
=
\sum_{j\neq k}a_{k,j}p_j^{(r)}+\nu_k .
\label{eq:Ik_current}
\end{equation}

Since \(\log_2(\cdot)\) is concave, its first-order Taylor expansion at \(\Omega_k^{(r)}\) provides a global upper bound, $
\log_2(\Omega_k) \leq \widetilde{\Psi}_k,
$ where
\begin{equation}
\widetilde{\Psi}_k
=
\log_2 \Omega_k^{(r)}
+
\sum_{j\neq k}
\frac{
a_{k,j}\left(p_j-p_j^{(r)}\right)
}{
\Omega_k^{(r)}\ln 2
}.
\label{eq:Psi_tilde}
\end{equation}
\vspace{-0.5em}
By substituting the upper bound \eqref{eq:Psi_tilde} into \eqref{eq:rate_dc}, we obtain the following concave lower bound of \(R_k\):
\begin{equation}
\widetilde{R}_k
=
B\log_2
\left(
\sum_{j\in\mathcal{K}}a_{k,j}p_j+\nu_k
\right)
-
B\widetilde{\Psi}_k.
\label{eq:rate_lower_bound}
\end{equation}
This lower bound is tight at $\bm{p}=\bm{p}^{(r)}$ and satisfies $
\widetilde{R}_k
\leq
R_k.
$
Therefore, replacing $R_k$ with $\widetilde{R}_k$ yields a conservative convex approximation of the transmit power update. 
The transmit power subproblem is then formulated as
\begin{subequations}\label{prob:power_subproblem}
\begin{align}
(\mathcal{P}3):\max_{\bm{p},\tau} \quad
& \tau \tag{\theequation}\label{prob:power_subproblem_obj}\\
\mathrm{s.t.}\quad
& \frac{\widetilde{R}_k}{\rho_k}
-
\lambda P_k^{\mathrm{tot}}
\geq \tau,
\quad \forall k\in\mathcal{K}, \label{prob:power_subproblem_a}\\
&\eqref{prob:P1a}\text{--}\eqref{prob:P1b}. \label{prob:power_subproblem_b}
\end{align}
\end{subequations}
Since $\widetilde{R}_k$ is concave with respect to $\bm{p}$ and $P_k^{\mathrm{ctrl}}$ is affine in $p_k$ for fixed $\bm{\rho}$ and $\bm{d}$, problem \((\mathcal{P}3)\) is convex and can be efficiently solved using standard convex solvers.

\vspace{-0.5em}
\subsection{Movement Distance Update}
For fixed $\bm{p}$ and $\bm{\rho}$, the movement distance affects the channel gain through the distance-dependent path loss, making the update of $\bm{d}$ non-convex. Therefore, we adopt a coordinate-wise grid search. Specifically, for each agent $k$, $d_k$ is searched over $d_k \in [d_k^{\min}, d_k^{\max}]$.

With the movement distances of the other agents fixed, each candidate value of $d_k$ is evaluated by recalculating the achievable rates and the corresponding max-min subtractive objective. The movement distance of agent $k$ is updated as
\begin{equation}
d_k^{\star}
=
\arg\max_{d_k\in[d_k^{\min},d_k^{\max}]}
\min_{i\in\mathcal{K}}
\left(
R_i^{\mathrm{eff}}
-
\lambda P_i^{\mathrm{tot}}
\right).
\label{eq:d_update}
\end{equation}
The candidate that gives the largest objective value while satisfying the controllable power constraint \eqref{prob:P1a} is selected.

\begin{algorithm}[t]
\caption{Proposed AO-Dinkelbach Algorithm}
\label{alg:proposed}
\begin{algorithmic}[1]
\State Initialize feasible $\bm{p}^{(0)}$, $\bm{d}^{(0)}$, and $\bm{\rho}^{(0)}$.
\State Initialize $\lambda^{(0)}=\min_{k\in\mathcal{K}}\eta_k^{(0)}$ and set $n=0$.

\Repeat
    \State Set $r=0$.

    \Repeat
        \State Update $\{\bm{w}_k\}$ using the MMSE rule in \eqref{eq:mmse}.
        \State Update $\bm{\rho}$ by solving \eqref{rho_update}.
        \State Update $\bm{p}$ by solving the SCA-based problem \eqref{prob:power_subproblem_obj}.
        \State Update $\bm{d}$ via coordinate-wise grid search in \eqref{eq:d_update}.
        \State $r\gets r+1$.
    \Until{inner-loop convergence}
    \State Update $\lambda^{(n+1)}$ according to \eqref{lamada}.
    \State $n\gets n+1$.
\Until{Dinkelbach convergence}
\State \Return $(\bm{p}^{\star},\bm{d}^{\star},\bm{\rho}^{\star})$.
\end{algorithmic}
\end{algorithm}
\subsection{Overall Algorithm}
Based on the above subproblem updates, the overall AO-Dinkelbach procedure is summarized in Algorithm~\ref{alg:proposed}. 
Here, $n$ denotes the outer Dinkelbach iteration and $r$ denotes the inner AO iteration. 
For each fixed $\lambda^{(n)}$, the algorithm updates the MMSE receive combiners and then alternately optimizes $\bm{\rho}$, $\bm{p}$, and $\bm{d}$ until the inner loop converges. 
The Dinkelbach parameter is then updated according to \eqref{lamada}, and the procedure repeats until the convergence condition is satisfied.


\subsection{Complexity Analysis}

Let \(I_{\mathrm{out}}\) and \(I_{\mathrm{in}}\) denote the numbers of outer Dinkelbach iterations and inner AO iterations, respectively. In each inner iteration, the MMSE combiner update is dominated by the inversion of an \(M\times M\) matrix. The SCA-based power update is solved by a standard convex solver with polynomial complexity in the number of agents \(K\). The coordinate-wise movement update evaluates \(K N_d\) candidate points, where \(N_d\) is the number of grid points for each agent, and each candidate evaluation requires channel and rate recalculation. Therefore, the overall computational complexity scales linearly with \(I_{\mathrm{out}}\), \(I_{\mathrm{in}}\), and \(N_d\), and polynomially with \(K\) and \(M\).


\section{Simulation Results}
\label{sec:results}
In this section, we evaluate the proposed joint optimization algorithm through simulations. We first present the setup and parameters, followed by comparisons with the benchmarks.
\vspace{-0.6em}
\subsection{Simulation Setup}
We consider an uplink multi-agent semantic communication system, where $K$ mobile agents simultaneously transmit data to a BS equipped with $M$ antennas. The wireless channel includes distance-dependent large-scale path loss and small-scale Rayleigh fading, with the large-scale channel gain modeled as \(\beta_k(d_k)=\beta_0(D_k-d_k)^{-\delta}\). Unless otherwise specified, the simulation parameters are summarized in Table~I. For Fig.~3, the controllable power budget is fixed at \(P_k^{\max}=23\) dBm, and a normalized distance factor \(\alpha\) is applied to scale the initial distances as \(D_k=\alpha D_k^0\), where \(D_k^0\) denotes the nominal distance in Table~I. For Fig.~4, the initial distances are set as \([120,130,140,150,160,170,180]\) m, the first \(K\) entries are used when the number of agents \(K\) changes, and the controllable power budget is fixed at \(P_k^{\max}=28\) dBm.


\begin{table}[t]
\centering
\caption{Simulation parameters}
\label{tab:sim_params}
\renewcommand{\arraystretch}{1.0}
\footnotesize
\begin{tabular}{l l}
\hline
\textbf{Parameters} & \textbf{Value} \\
\hline
$K$: number of agents & $4$ \\
$M$: number of BS antennas & $8$ \\
$B$: system bandwidth & $10$ MHz \\
$N_0$: noise power spectral density & $-174$ dBm/Hz \\
$\beta_0$: reference channel gain at $1$ m & $-40$ dB \\
$\delta$: path-loss exponent & $3.5$ \\
$D_k$: initial distance of agent $k$ & $[120,150,130,180]$ m \\
$P_k^{\mathrm{c}}$: fixed circuit power & $0.5$ W \\
$A_k$: semantic compression coefficient & $0.2$ W\\
$F_k$: mobility energy coefficient & $2\times10^{-3}$ J/m \\
$T_k$: movement duration & $10$ s \\
$d_k^{\max}$: maximum movement distance & $30$ m \\
$\rho_k$: compression ratio range & $[0.1,1]$ \\
$P_k^{\max}$: controllable power budget & $20$--$36$ dBm \\
$N_d$: number of grid points for $d_k$ & $45$ \\
\hline
\end{tabular}
\vspace{-2em}
\end{table}

To evaluate the effectiveness of the proposed joint optimization scheme, we compare it with two baseline schemes: 
\begin{itemize}
    \item Baseline 1 (no-mobility): agents optimize transmit power and compression ratios but remain at their initial locations ($d_k = 0$).
    \item Baseline 2 (no-compression): agents perform conventional communication without semantic compression ($\rho_k = 1$) but optimize transmit power and movement distances.
\end{itemize}
\vspace{-0.6em}

\subsection{Performance Analysis}
\begin{figure}[t]
    \centering
    \includegraphics[width=0.85\linewidth]{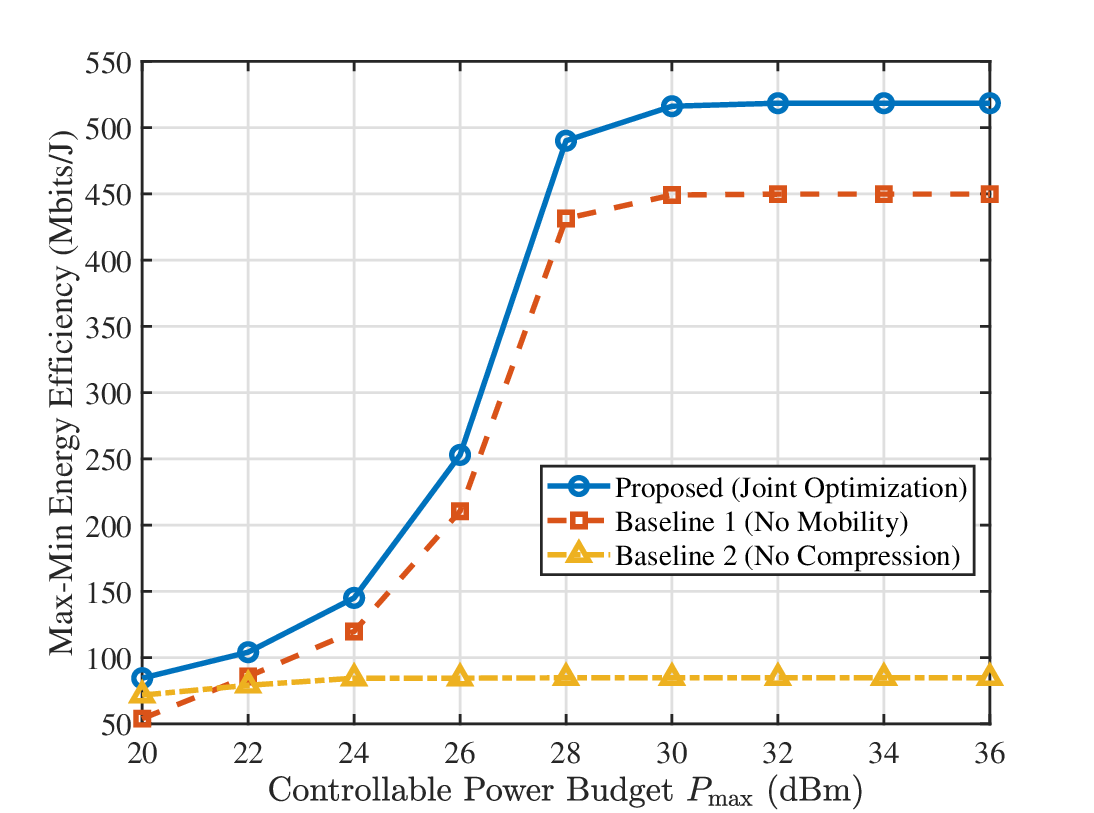} 
    \vspace{-1em}
    \caption{Max-min EE as a function of controllable power budget.}
    \label{fig:power}
    \vspace{-1.5em}
\end{figure}

Fig.~2 illustrates the max-min EE versus the controllable power budget $P_{\max}$. 
The proposed scheme consistently achieves the highest EE across the whole power range. 
When $P_{\max}$ increases from 20 dBm to 28 dBm, the EE of the proposed scheme rises from approximately 85 Mbits/J to 490 Mbits/J, while no-mobility increases from about 55 Mbits/J to 430 Mbits/J and no-compression only improves from around 73 Mbits/J to 85 Mbits/J. 
This shows that semantic compression significantly improves EE by reducing the effective transmission load, while mobility control further enhances the channel condition. 
As $P_{\max}$ further increases, the proposed scheme gradually saturates at around 515 Mbits/J, whereas no-mobility saturates at about 450 Mbits/J and no-compression remains nearly flat. 
This indicates that, at high power budgets, the system performance is increasingly limited by rate saturation and multi-user interference, so additional controllable power brings only marginal EE improvement. 
Overall, the proposed scheme provides about 60--70 Mbits/J gain over no-mobility in the medium-to-high power region and achieves more than fivefold improvement over no-compression at high power budgets.


\begin{figure}[t]
    \centering    \includegraphics[width=0.85\linewidth]{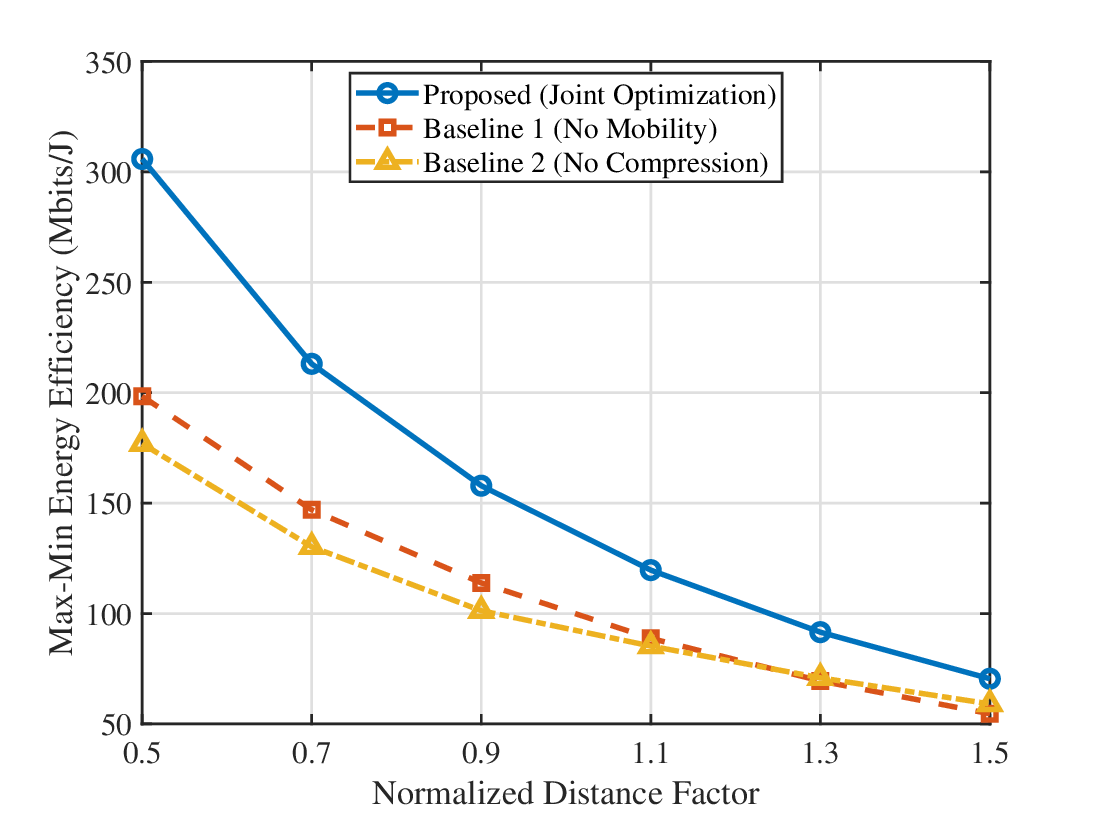}
    \vspace{-0.6em}
    \caption{Max-min EE as a function of normalized distance scaling factor.}
    \label{fig:distance}
    \vspace{-1.5em}
\end{figure}

As shown in Fig.~\ref{fig:distance}, the max-min EE decreases monotonically as the normalized distance scaling factor increases for all schemes. 
Specifically, when the normalized distance factor increases from 0.5 to 1.5, the EE of the proposed scheme decreases from approximately 305~Mbits/J to about 70~Mbits/J. 
This trend is mainly caused by the increased path loss associated with larger communication distances, which weakens the received signal strength and reduces the achievable uplink rate under the same power budget. 
Despite this degradation, the proposed scheme consistently outperforms both baselines across the entire distance range. 
For example, at a distance factor of 0.5, the proposed scheme achieves around 305~Mbits/J, compared with about 198~Mbits/J for no-mobility and 178~Mbits/J for no-compression. 
At a distance factor of 1.1, the proposed scheme still achieves approximately 120~Mbits/J, while no-mobility and no-compression decrease to around 88~Mbits/J and 85~Mbits/J, respectively. 
These results indicate that mobility control helps improve channel conditions, while semantic compression reduces the effective transmission load. 
Therefore, the proposed joint design provides better resilience against distance-induced path loss.

\begin{figure}[t]
    \centering
    \includegraphics[width=0.85\linewidth]{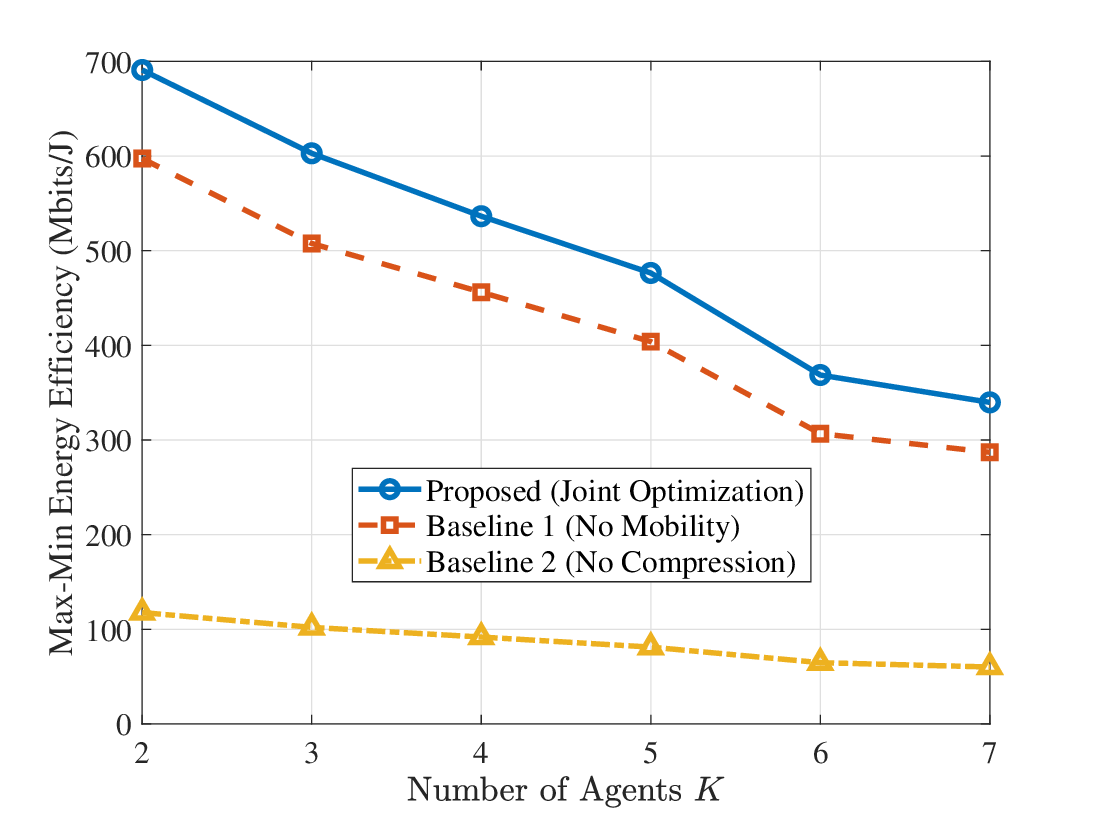}
    \vspace{-0.7em}
    \caption{Max-min EE as a function of the number of agents $K$.}
    \label{fig:users}
    \vspace{-1.8em}
\end{figure}

Fig.~\ref{fig:users} illustrates the max-min EE versus the number of agents $K$, where the results are averaged over 100 independent channel realizations. 
As $K$ increases, the max-min EE of all schemes decreases due to stronger multi-user interference and stricter max-min fairness requirements. 
Nevertheless, the proposed joint optimization scheme consistently achieves the best performance across all tested network sizes. 
For example, when $K=5$, the proposed scheme achieves about 480~Mbits/J, compared with approximately 405~Mbits/J for no-mobility and 80~Mbits/J for no-compression. 
This corresponds to an EE improvement of about 18.5\% over no-mobility and a much larger gain over no-compression. 
The results demonstrate that jointly optimizing transmit power, semantic compression, and mobility improves the scalability of the system, while the large gap between the proposed scheme and no-compression confirms the critical role of semantic compression in enhancing energy efficiency.


\section{Conclusion}
\label{sec:conclusion}
In this paper, we proposed a communication-computation-mobility framework for uplink MEAN, where semantic compression ratio, transmit power, and movement distance are jointly optimized to improve max-min EE. We formulated the resulting problem as a non-convex fractional program and developed an AO-Dinkelbach algorithm with SCA-based power optimization and grid-search-based movement update. Simulation results showed that the proposed scheme outperforms no-mobility and no-compression baselines. Future work will extend the framework to multi-cell MEAN with imperfect channel state information and task-oriented semantic accuracy constraints.



\bibliographystyle{ieeetr}
\def\baselinestretch{0.92}
\bibliography{ref}
\end{document}